\documentclass[%
 reprint,
 amsmath,amssymb,
 onecolumn,
prb,
]{revtex4-2}

\usepackage{graphicx}% Include figure files
\usepackage{dcolumn}% Align table columns on decimal point
\usepackage{bm}% bold math
\usepackage{physics}

\begin{document}

\preprint{}

\title{Mixtures of dipolar gases in two dimensions: a quantum Monte Carlo 
study}

\author{Sergi Pradas}
 %\altaffiliation[Also at ]{Physics Department, XYZ University.}%Lines break 
%automatically or can be forced with \\
\author{Jordi Boronat}%
\affiliation{%
Departament de Física, Campus Nord B4-B5, Universitat Politècnica de Catalunya,
E-08034 Barcelona, Spain
}%

\date{\today}% It is always \today, today,

\begin{abstract}
We studied the miscibility of two dipolar quantum gases in the limit
of zero temperature. The system under study is composed by a mixture of two
Bose gases with dominant dipolar interaction in a two-dimensional harmonic
confinement. The dipolar moments are considered all to be perpendicular to the
plane, turning the dipolar potential in a purely repulsive and isotropic model.
Our analysis is carried out by using the diffusion Monte Carlo method  which
allows for an exact solution to the many-body problem within some statistical
noise. Our results show that the miscibility between the two species is rather
constrained as a function of the relative dipolar moments and masses of the two
components. A narrow regime is predicted where both species mix and we
introduce an adimensional parameter whose value predicts quite accurately the
miscibility of the two dipolar gases.
\end{abstract}

\maketitle

%%%%%%%%%%%%%%%%%%%%%%%%%%%%%%%%%%%%%%%%%%

\section{Introduction} \label{sec:introduction}

Ultracold Bose and Fermi gases have proved to be the best platform for the
study of quantum many-body systems~\cite{Pethick}. Their versatility and fine tuning of the
interatomic interactions allow for the study of many phenomena which are
difficult to attain in real systems. They offer the opportunity of using a 
laboratory
as a quantum simulator of Hamiltonians proposed from theory which could be
difficult to manage using classical computers and algorithms~\cite{Georgescu,Altman}. Many of the atoms
used to achieve the Bose-Einstein condensate (BEC) state interact among
themselves with a contact short-range potential, that depends only on the 
$s$-wave
scattering length, due to the extreme low density of these gases. However, it
has been also possible to cool down to degeneracy gases composed by atoms with
a permanent magnetic moment~\cite{lahaye2009physics}. Exploiting Feshbach 
resonances, it is proved that
the dominant interaction is no more the contact potential but the dipolar
interaction between the atomic magnetic moments~\cite{Koch}. The dipolar potential decays as 
$r^{-3}$ and thus the interaction effects become fundamental in the properties 
of the gas.

First experiments showing dipolar effects were carried out with Cr, with a
magnetic moment $\mu=6 \mu_B$~\cite{Koch}. In 
the last years, two more candidates have joined to this class of materials, 
Dy~\cite{lu2011strongly} with
$\mu=10 \mu_B$ and Er~\cite{aikawa2012bose} with $\mu=7 \mu_B$, widening 
significantly the possibilities
for observing the two main features of these systems: anisotropy and
slow-decaying two-body interactions~\cite{Norcia}. The different interaction 
between
side-by-side moments (repulsive) and head-to-tail ones (attractive) leads to
the formation of self-bound liquid drops if the number of atoms is above a
threshold known as critical atom number~\cite{Ferrier-Barbut}. By changing the total scattering
length of the system, one can see how the critical atom number increases when
the scattering length also increases~\cite{bottcher2019dilute}. Under  
proper harmonic confinement,
and when the number of atoms is large enough, one observes that the system
arranges in drops forming a linear array, if the trapping is cigar shaped, and a
triangular one in a plane, if it has the pancake form. Interestingly, these
patterns emulate a crystal but where every site is not monoatomic but occupied
by a multiparticle drop. Recent experimental work claims that these solid-like
patterns show coherence and thus are examples of the pursued supersolid state
of matter~\cite{Tanzi,Bottcher,Chomaz}.

The field of ultracold dipolar gases has entered on an even more rich landscape
with the realization of Er-Dy 
mixtures~\cite{Ravensbergen,trautmann2018dipolar,Politi}. The interplay between 
the two dipolar
species opens new scenarios like the formation of mixed dipolar drops and the
possible stability of mixed supersolids, with arrays composed by
single-species drops or mixed drops. A key ingredient in this discussion is the
miscibility of the two species since inmiscibility would hinder the observation
of these new intriguing phases. Recent measures on this system show that both
components tend to be phase separated, both due to the gravitational sag
originated by the different masses of Er and Dy and by an overall repulsive
interaction between both condensates~\cite{Politi}.

In the present work, we use the ab-initio diffusion Monte Carlo (DMC) method to 
study a mixture of two dipolar Bose gases harmonically confined in two 
dimensions (2D). In our analysis, we assume that all the dipoles are oriented 
perpendicularly to the plane and thus interact with a fully repulsive $1/r^3$ 
potential. A single dipolar gas in the same conditions was studied some 
time ago with DMC but in an extended configuration, free from confinement. It 
was shown that the gas becomes a triangular crystal when the density 
increases~\cite{Astradipol}. 
If the dipoles are not perpendicular to the plane but tilted a certain angle, 
the interaction becomes anisotropic and, beyond a certain critical angle, it 
collapses. That anisotropy produces a rich diagram, with a stable stripe 
phase~\cite{Bombin1}, 
which is indeed a supersolid or superstripe that suffers a 
Berezinskii-Kostrelitz-Thouless phase transition at finite 
temperature~\cite{Bombin2}. 

As a function of the ratio between both the dipolar moments and the masses of 
the two species in the mixture, we analyze the miscibility of the two gases. 
Our results show that both species are miscible only in a restricted area in 
the dipolar moment, mass  ratio plane where both ratios are close to one. In 
the 
majority of situations that we analyzed, we observe that the two confined gases 
do not mix: one species remains in the center and the second goes to the 
surface. 
If the trap is deformed, we observe that in some cases the external component 
appears in two separated blobs, separated by the inner species. Finally, we 
particularize our study to the Er-Dy mixture and predict that both species do 
not mix, in agreement with available experimental data.

The rest of the paper is organized as follows. In Sec. \ref{sec:materials_and_methods}, we discuss the 
quantum Monte Carlo methods used in our study and the miscibility criterion 
that accounts well for the phase diagram. In Sec. \ref{sec:results}, we present the results 
obtained for both isotropic and anisotropic traps and analyze the particular 
case of an Er-Dy mixture, which is the one observed recently in experiments. 
Finally, Sec. \ref{sec:Discussion} comprises the summary of the main results and the conclusions 
of our work.

%%%%%%%%%%%%%%%%%%%%%%%%%%%%%%%%%%%%%%%%%%
\section{Quantum Monte  Carlo Methods} \label{sec:materials_and_methods}

\subsection{Hamiltonian} \label{subsec:mixtures_under_study}

The object of study of this work are two-component dipolar bosonic mixtures at 
zero temperature, in a purely two dimensional geometry. We consider that 
all the magnetic moments are perpendicular to the plane and the gas is confined 
 by either an isotropic harmonic trap or an anisotropic one. The dipole-dipole 
interaction (DDI) potential between two identical particles is given by
\begin{equation}
    V_{dd}(\mathbf{r}) = 
\frac{C_{dd}}{4\pi}\, \frac{\hat{\mathbf{p}}_{1}\cdot\hat{\mathbf{p}}_{2} 
-3(\hat{\mathbf{p}}_{1}\cdot\hat{\mathbf{r}})(\hat{\mathbf{p}}_{2}\cdot\hat{
\mathbf{r}})}{r^3} \ ,
\end{equation} 
with  $C_{dd} = \mu_0\mu^2$, $\mu_0$ being the magnetic 
permeability of free space and $\mu$ the particle's magnetic moment. 
$\hat{\mathbf{p}}_1$ and $\hat{\mathbf{p}}_2$ are the vectors pointing in the 
direction of the dipole's moment of particles $1$ and $2$, respectively, and 
$\hat{\mathbf{r}} = \mathbf{r}/r$ is the unit position vector. In our 
case, with the particles confined to the $xy$ plane and polarized parallel to 
the $z$ axis, the interaction is always isotropic and repulsive,
\begin{equation}
    V_{dd}(\mathbf{r}) = V_{dd}(r) = \frac{C_{dd}}{4\pi}\frac{1}{r^3}.
\end{equation}
The mixture is confined by an harmonic oscillator (ho) potential, that in the 
isotropic case is given by
\begin{equation}
    V_{ho}(r_i) = \frac{1}{2}m\omega^{2}r_{i}^{2} \ ,
\end{equation}
where $m$ is the particle's mass, $\omega$ is the ho's trapping frequency, and
$r_{i}$ is the particle's distance to the origin, which matches the center of 
the trap. The full Hamiltonian of the system is then given by
\begin{align}
    H(\mathbf{R}) = -\frac{\hbar^2}{2m_{1}}\nabla_{\mathbf{R}_{1}}^{2} \;\; -\frac{\hbar^2}{2m_{2}}\nabla_{\mathbf{R}_{2}}^{2} \;\; + \;\; \frac{1}{2}m_{1}\omega_{1}^{2}\sum_{i=1}^{N_{1}}r_{i}^{2} \;\; + \;\;
    \frac{1}{2}m_{2}\omega_{2}^{2}\sum_{k=N_{1}+1}^{N}r_{k}^{2} \;\; + \nonumber \\
    + \; \; 
\frac{\mu_0\mu_{1}^{2}}{4\pi}\sum_{i=1}^{N_{1}-1}\sum_{j=i+1}^{N_{1}}\frac{1}{r_
{ij}^{3}} \; \; + \;\; 
\frac{\mu_0\mu_{2}^{2}}{4\pi}\sum_{i=N_{1}+1}^{N-1}\sum_{j=i+1}^{N}\frac{1}{r_{
ij}^{3}} \;\; + \;\; 
\frac{\mu_0\mu_{1}\mu_{2}}{4\pi}\sum_{i=1}^{N_{1}}\sum_{j=N_{1}+1}^{N}\frac{1}{
r_{ij}^{3}} \ ,
\label{hamiltonian1}
\end{align}
with $N = N_1 + N_2$, $N_1$ and $N_2$ being the number of particles of each 
type in our mixture, and $m_1$ and $m_2$ the particle's masses. In order to 
simplify Eq. (\ref{hamiltonian1}), we have used $\mathbf{R} = 
\{\mathbf{r}_1,\mathbf{r}_2,\cdots,\mathbf{r}_N\}$ as the whole coordinate set 
such that $\sum_{i}\nabla_{i}^{2} = \nabla_{\mathbf{R}}^{2}$.
Finally, $\omega_1$ and $\omega_2$ are the trapping frequencies and $\mu_1$ and 
$\mu_2$ are the magnetic dipole moments of type 1 and 2 particles, respectively, 
with $\mathbf{r}_{ij} = \mathbf{r}_i - \mathbf{r}_j$. 

As in previous studies~\cite{Astradipol}, we use dipolar units (for 
species 1),
\begin{equation}
    r_0 = \frac{m_{1}C_{dd}^{(11)}}{4\pi\hbar^{2}} = 
\frac{m_{1}\mu_{0}\mu_{1}^{2}}{4\pi\hbar^2} \quad\mathrm{\&}\quad E_0 = 
\frac{\hbar^2}{m_{1}r_{0}^{2}} \ ,
\end{equation}
with $r_0$ and $E_0$ the units of distance and energy, respectively. Then, in 
these units the Hamiltonian is written as
\begin{align}
        H(\mathbf{R}^{*}) = -\frac{1}{2}\nabla_{\mathbf{R}_{1}^{*}}^{2} \;\; 
-\frac{1}{2}\frac{m_1}{m_{2}}\nabla_{\mathbf{R}_{2}^{*}}^{2} \;\; + \;\; 
\frac{1}{2}A_{1}\sum_{i=1}^{N_{1}}r_{i}^{*,2} \;\; + \;\;
    \frac{1}{2}A_{2}\sum_{k=N_{1}+1}^{N}r_{k}^{*,2} \;\; + \nonumber \\
    + \; \; \sum_{i=1}^{N_{1}}\sum_{j=i+1}^{N_{1}-1}\frac{1}{r_{ij}^{*,3}} \; \; 
+ \;\; 
\left(\frac{\mu_2}{\mu_1}\right)^{2}\sum_{i=N_{1}+1}^{N-1}\sum_{j=i+1}^{N}\frac{
1}{r_{ij}^{*,3}} \;\; + \;\; 
\frac{\mu_{2}}{\mu_1}\sum_{i=1}^{N_{1}}\sum_{j=N_{1}+1}^{N}\frac{1}{r_{ij}^{*,3}
} \ ,
\label{eq:hamil_iso}
\end{align}
where the superscript '${}^{*}$' denotes the use of normalized units $r^{*} = 
r/r_0$. The strength of the harmonic confinement in both species is 
\begin{equation}
    A_1 = \frac{m_{1}^{2}\omega_{1}^{2}r_{0}^{4}}{\hbar^{2}} = \left(\frac{r_0}{l_1}\right)^4
    \quad\mathrm{\&}\quad
    A_2 = 
\frac{m_{1}m_{2}\omega_{2}^{2}r_{0}^{4}}{\hbar^{2}}=\frac{m_1}{m_2}\left(\frac{
r_0}{l_2}\right)^{4}=\frac{m_{2}\omega_{2}^{2}}{m_{1}\omega_{1}^{2}}A_1 \ , 
\label{eq:A_def}
\end{equation}
with $l_{1} = \sqrt{\hbar/m_{1}\omega_{1}}$ and $l_{2} = 
\sqrt{\hbar/m_{2}\omega_{2}}$ the harmonic oscillator lengths. 

We have also explored the effects induced by an anisotropic confinement. The 
Hamiltonian in this case is
\begin{align}
        H(\mathbf{R}^{*}) = -\frac{1}{2}\nabla_{\mathbf{R}_{1}^{*}}^{2} \: -\frac{1}{2}\frac{m_1}{m_{2}}\nabla_{\mathbf{R}_{2}^{*}}^{2} \: + \: \frac{A_{x,1}}{2}\sum_{i=1}^{N_{1}}x_{i}^{*,2} \: + \:
    \frac{A_{x,2}}{2}\sum_{k=N_{1}+1}^{N}x_{k}^{*,2} \: + \:\frac{A_{y,1}}{2}\sum_{i=1}^{N_{1}}y_{i}^{*,2} \: \nonumber \\
    + \: \frac{A_{y,2}}{2}\sum_{k=N_{1}+1}^{N}y_{k}^{*,2} \: +  \: 
\sum_{i=1}^{N_{1}-1}\sum_{j=i+1}^{N_{1}}\frac{1}{r_{ij}^{*,3}} \: + \: 
\left(\frac{\mu_2}{\mu_1}\right)^{2}\sum_{i=N_{1}+1}^{N-1}\sum_{j=i+1}^{N}\frac{
1}{r_{ij}^{*,3}} \: + \: 
\frac{\mu_{2}}{\mu_1}\sum_{i=1}^{N_{1}}\sum_{j=N_{1}+1}^{N}\frac{1}{r_{ij}^{*,3}
} \ ,
\label{eq:hamil_aniso}
\end{align}
with
\begin{align}
    A_{x,1} &= \frac{m_{1}^{2}\omega_{x,1}^{2}r_{0}^{4}}{\hbar^{2}} = \left(\frac{r_0}{l_{x,1}}\right)^4 \text{ , \:}
    A_{x,2} &= \frac{m_{1}m_{2}\omega_{x,2}^{2}r_{0}^{4}}{\hbar^{2}}=\frac{m_1}{m_2}\left(\frac{r_0}{l_{x,2}}\right)^{4}=\frac{m_{2}\omega_{x,2}^{2}}{m_{1}\omega_{x,1}^{2}}A_{x,1} \nonumber \\
    A_{y,1} &= \frac{m_{1}^{2}\omega_{y,1}^{2}r_{0}^{4}}{\hbar^{2}} = \left(\frac{r_0}{l_{y,1}}\right)^4 \text{ , \:}
    A_{y,2} &= 
\frac{m_{1}m_{2}\omega_{y,2}^{2}r_{0}^{4}}{\hbar^{2}}=\frac{m_1}{m_2}\left(\frac
{r_0}{l_{y,2}}\right)^{4}=\frac{m_{2}\omega_{y,2}^{2}}{m_{1}\omega_{y,1}^{2}}A_{
y,1} \ .  
\label{eq:A_aniso_def}
\end{align}
The oscillator lengths are now: $l_{x,1} = \sqrt{\hbar/m_{1}\omega_{x,1}}$, 
$l_{x,2} = \sqrt{\hbar/m_{2}\omega_{x,2}}$, $l_{y,1} = 
\sqrt{\hbar/m_{1}\omega_{y,1}}$, and $l_{y,2} = 
\sqrt{\hbar/m_{2}\omega_{y,2}}$. 

To reduce the number of variables of our numerical simulations,
 we assume that 
both types of particles are under the presence of the same harmonic potential, 
with strengths $A_x$ in the $x$-direction and $A_y$ in the $y$-direction. 
Therefore,  $m_1\omega_{x,1}^2 = 
m_2\omega_{x,2}^2$ and $m_1\omega_{y,1}^2 = m_2\omega_{y,2}^2$, 
$\omega_{x,\alpha}$ and $\omega_{y,\alpha}$ being the confinement frequencies 
for type $\alpha$ particles along the $x$ and $y$ axes, respectively. This also 
applies to the isotropic case, where we consider $A_1= A_2$ via the use of 
different confinement frequencies verifying $m_{1}\omega_{1}^2 = 
m_{2}\omega_{2}^2$.

\subsection{Diffusion Monte Carlo} \label{subsec:DMC}

The main theoretical tool used in this work is the  diffusion Monte Carlo 
(DMC) method, which finds the ground-state energy of a many-particle system by 
propagating the imaginary-time Schrödinger equation exploiting the 
similarities with a diffusion process. 

The  time-dependent Schrödinger equation associated to a 
system of $N$ particles, written in imaginary time $\tau = it/\hbar$, is
\begin{equation}
-\frac{\partial \ket{\Psi(\tau)}}{\partial \tau} = (\hat{H} - E_T)\ket{\Psi(\tau)}. \label{eq:TDSEformal}
\end{equation}
Its solution is given by $\ket{\Psi(\tau)} = \hat{U}(\tau,0)\ket{\Psi(0)}$, 
where $\hat{U}(\tau,0)=\text{exp}[-(\hat{H}-E_T)\tau)]$ is the imaginary-time 
evolution operator and $E_T$ a reference energy. Projecting in coordinate 
space,
\begin{equation}
\Psi(\mathbf{R},\tau) = \int d\mathbf{R'}\, 
G(\mathbf{R},\mathbf{R'},\tau)\Psi(\mathbf{R},0) \ ,
\label{Green}
\end{equation}
with $\bra{\mathbf{R}}\hat{U}(\tau,0)\ket{\mathbf{R'}} \equiv
G(\mathbf{R},\mathbf{R'},\tau)$ the Green function and 
$G(\mathbf{R},\mathbf{R'},0) = \delta(\mathbf{R}-\mathbf{R'})$ its initial 
condition. The above integral (\ref{Green})is in principle intractable  
due to the 
non-commutativity of the kinetic and potential operators that appear in the 
Green function. As we will shortly discuss, we may avoid this by computing 
short-time approximations for $G(\mathbf{R},\mathbf{R'},\tau)$ and using the 
convolution property of Eq. (\ref{Green}). 

As in many Monte Carlo simulations, it is convenient to introduce 
importance sampling to reduce the variance to a manageable level. In DMC, this 
is carried out by solving the Schrödinger equation for the wave function
\begin{equation}
f(\mathbf{R},\tau) \equiv \Psi(\mathbf{R},\tau)\Psi_{T}(\mathbf{R}), \label{eq:f-psi_t}
\end{equation}
where $\Psi_{T}(\mathbf{R})$ is a time-independent trial wave function, which in 
our case is previously optimized using the variational Monte Carlo (VMC) 
method~\cite{guardiola1998monte}, and whose specific form for the present 
problem will be discussed  in Section~\ref{subsec:trial}. $\Psi_T$ acts as a 
guiding wave function which drives the system away from regions of the 
phase space where the interatomic potential is strongly repulsive, or even 
divergent, 
 and increases the 
sampling of regions where the wave function is expected to be large. With that, 
the Schrödinger equation may be rewritten as
\begin{equation}
-\frac{\partial f(\mathbf{R},\tau)}{\partial \tau} = (\hat{O}_{K} + \hat{O}_{D} 
+ \hat{O}_{B})f(\mathbf{R},\tau) \ , 
\label{eq:DMCformal}
\end{equation}
where
\begin{equation}
\hat{O}_{K} = -D\,\nabla_{R}^{2} \;, \;\;\; \hat{O}_{D} = D\left[(\nabla_{R}\cdot\mathbf{F}(\mathbf{R})) + \mathbf{F}(\mathbf{R})\cdot\nabla_{\mathbf{R}} \right] \;\;\; \& \;\;\; \hat{O}_{B} = E_L(\mathbf{R}) - E_T
\end{equation}
are known as the kinetic, drift, and branching operators, respectively. 
$D=\hbar^2/(2m)$  and 
$E_{L}(\mathbf{R}) = 
\frac{H(\mathbf{R})\Psi_{T}(\mathbf{R})}{\Psi_{T}(\mathbf{R})}$ is the local 
energy. The kinetic term, which gives DMC its name, 
is the same as a classical diffusion operator, $\hat{O}_D$ guides the diffusion 
process, and, as we will see, $\hat{O}_B$ promotes lower energy configurations 
and removes higher energy ones. Then, in a formal sense, this leads to
\begin{equation}
f(\mathbf{R},\tau) = \int d\mathbf{R'}\, G(\mathbf{R},\mathbf{R'},\tau)f(\mathbf{R},0),
\end{equation}
with $G(\mathbf{R},\mathbf{R'},0) = \delta(\mathbf{R}-\mathbf{R'})$. 
Specifically, $G(\mathbf{R},\mathbf{R'},\tau) = \text{exp}[-(\hat{O}_K + 
\hat{O}_D + \hat{O}_B)\tau]$. The situation is formally the same as in the 
non-importance sampling case, as these three operators do not commute, and thus 
a short-time approximation needs to be used. In fact, the Green functions 
associated to each operator can be calculated in the limit $\tau \to 0$.

The kinetic part, $G_K(\mathbf{R},\mathbf{R'},\tau)$ is 
straightforwardly obtained by using the completeness relation of the momentum 
basis and gaussian integration, leading to
\begin{equation}
G_{K}(\mathbf{R},\mathbf{R'},\tau) = 
\bra{\mathbf{R}}e^{-\hat{O}_{K}\tau}\ket{\mathbf{R'}} = (4\pi D 
\tau)^{-dN/2}\text{exp}\left[-\frac{(\mathbf{R}-\mathbf{R'})^2}{4D\tau}\right] 
\ , 
\label{eq:diffusion_DMC}
\end{equation}
with  $d$ the dimension of the system. Concerning  the 
drift term, it  verifies \cite{vrbik1986quadratic}
\begin{equation}
G_D(\mathbf{R},\mathbf{R'},\tau) = 
\bra{\mathbf{R}}e^{-\hat{O}_{D}\tau}\ket{\mathbf{R'}} =  
\delta(\mathbf{R}-\mathcal{R}'(\tau)) \ ,
\end{equation}
where $\mathcal{R}'(\tau)$ is deterministically calculated by solving the 
differential equation,
\begin{equation}
\frac{d\mathcal{R}(\tau)}{d\tau} = D \, \mathbf{F}(\mathcal{R}(\tau)) \ , 
\label{eq:diffdrift}
\end{equation}
with the initial condition $\mathcal{R}(0) = \mathbf{R}$. Finally, the Green 
function associated to the branching term is easily obtained as 
\begin{equation}
G_{B}(\mathbf{R},\mathbf{R'},\tau) = \bra{\mathbf{R}}e^{-(E_L(\mathbf{R}) - 
E_T)\tau)}\ket{\mathbf{R'}} = 
e^{-(E_L(\mathbf{R})-E_T)\tau}\delta(\mathbf{R}-\mathbf{R'}) \ .
\end{equation}

One might then be tempted to approximate $G(\mathbf{R},\mathbf{R'},\Delta\tau)$ 
for small $\Delta\tau$ in terms of $G_K$, $G_D$ and $G_B$ linearly in 
$\Delta\tau$ as
\begin{equation}
G(\mathbf{R},\mathbf{R'},\Delta\tau) = \bra{\mathbf{R}}e^{-(\hat{O}_K + 
\hat{O}_D + \hat{O}_B)\Delta\tau}\ket{\mathbf{R'}} = 
\bra{\mathbf{R}}e^{-\hat{O}_{K}\Delta\tau}e^{-\hat{O}_{D}\Delta\tau}e^{-\hat{O}_
{B}\Delta\tau}\ket{\mathbf{R'}} + \mathcal{O}(\Delta\tau^2) \ .
\end{equation}
However, that would lead to a strong dependence with $\Delta\tau$ that would 
require to use very small time-steps. 
Instead, we use an approximation accurate to order $\Delta\tau^2$ given by
\begin{equation}
G(\mathbf{R},\mathbf{R'},\Delta\tau) = 
\bra{\mathbf{R}}e^{-\hat{O}_{B}\Delta\tau/2}e^{-\hat{O}_{D}\Delta\tau/2}e^{-\hat
{O}_{K}\Delta\tau}e^{-\hat{O}_{D}\Delta\tau/2}e^{-\hat{O}_{B}\Delta\tau/2}\ket{
\mathcal{R'}} + \mathcal{O}(\Delta\tau^3) \ , 
\label{eq:formalquadraticapprox}
\end{equation}
which gives name to the method known as quadratic diffusion Monte 
Carlo~\cite{casudmc,llibremeu}, the one used throughout this work.

Having discussed the formal concepts associated to the DMC method, let us then 
comment  how 
it is applied in practice. In order to evolve the system, one represents it 
using sets of coordinates 
$\{\mathbf{R}_1,\mathbf{R}_2,\cdots,\mathbf{R}_{N_w}\}$ known as 
\textit{walkers}, each one being a configuration of the $N$-particle system, 
which is to be propagated separately. One then makes use of the approximation 
(Eq. \ref{eq:formalquadraticapprox}), with each operator in this 
expression applied in the same order each time-step. 

The diffusion operator is straightforwardly applied, as one can use the 
Box-Muller algorithm to sample from the desired gaussian distribution. The 
drift operator $\text{exp}(-\hat{O}_{D}\Delta\tau/2)$  is 
deterministic, and requires solving Eq.~(\ref{eq:diffdrift}) exactly to order 
$\Delta\tau^{2}$. We do this by using the second-order Runge-Kutta method. 
 Finally, the branching operator kills or reproduces 
walkers based on the difference between their local energy $E_L$ and the 
reference energy  $E_T$. This way, we promote the configurations with the 
lowest energy and remove the ones with a higher energy. To implement the 
branching term,  the number of copies of a given walker is 
calculated as
\begin{equation}
N_{\text{sons}} = 
\left[\text{exp}\left\{-\left(\frac{E_L(\mathbf{R})+E_L(\mathbf{R}')}{2}
-E_T\right)\tau\right\} + \eta \right] \ , 
\label{eq:Nsons_formula}
\end{equation}
where $\eta$ is sampled from the uniform probability  
distribution $U[0,1)$, $[\cdots]$ denotes the integer part, and 
$\mathbf{R}$, $\mathbf{R}'$ are walkers at two successive times.

In a DMC simulation, one fixes the time-step $\Delta\tau$, the 
desired 
number of walkers $N_{w}$, and chooses a 
guiding wave function whose parameters are  previously optimized using the VMC 
method. Then, after each time-step, a new set of walkers is obtained and 
after every block $E_T$ is updated to be the average local energy of the 
previous set. Approaching  
the limit $\tau \to \infty$, we get  the ground-state energy $E_0$ as a mean of 
the local energies of the walkers set.
 It is important to remark, 
however, that this estimate is only exact in the limits  
$\Delta\tau \to 0$ and $1/N_{w} \to 0$. This way, different simulations 
approaching these two limits have to be made in order to obtain the result as an 
extrapolation to such limits. This is the method used throughout this work, and 
any result shown in Section~\ref{sec:results} has been obtained by an initial 
optimization of the trial wave function via VMC, followed by its use as a 
guiding function in DMC, and the check for the optimum $N_w$ and $\Delta\tau$ 
values. 

Finally, to conclude this discussion on the DMC method we briefly describe the 
estimation of observables. In a DMC simulation we sample the mixed wave function 
$f(\mathbf{R},\tau) = \Psi(\mathbf{R},\tau)\Psi_T(\mathbf{R})$ as $\tau \to 
\infty$, such that $f(\mathbf{R},\tau \to \infty) \to 
\phi_0(\mathbf{R})\Psi_{T}(\mathbf{R})$, where $\phi_0$ is the exact ground 
state wave function, in a process known as mixed estimation. In case 
the operator being estimated is the Hamiltonian or commutes with 
it  this leads to an 
exact estimation within some statistical errors. However, for other operators, 
such as the potential energy or the density profile, the estimation is 
biased in a way difficult to assess \textit{a priori}. The simplest way to 
approximately correct this bias is to use an  extrapolated 
estimator, which makes use of both mixed (DMC) and variational (VMC) results. 
This result, however, is still biased in an unknown way as it still depends on 
the trial wave function. The solution to definitely eliminate that bias is by 
means of the pure estimation,  a technique based on forward 
walking, which we use   throughout this work as implemented in  
Ref.~\cite{casulleras1995unbiased}.

\subsection{The trial wave functions} \label{subsec:trial}

As we have discussed in the previous Section, a reasonable choice 
for the ground-state wave function is necessary to guide the diffusion process 
in DMC. The usual approach, and the one taken in this work, is to 
use the VMC method to optimize the trial wave function 
before using it in the DMC method.

Our system is a bosonic one, so the wave function must be 
symmetric with respect to exchange of particles. We include one- and two-body 
correlation factors in the usual Bijl-Jastrow form~\cite{jastrow1955many}, 
\begin{equation}
\Psi (\mathbf{R}) =   \psi_1 (\mathbf{R}) 
\, \psi_2 (\mathbf{R}) \ ,
\end{equation}
with
\begin{equation}
\psi_1 (\mathbf{R}) = \exp \left( \sum_{i=1}^{N_1} u_1^{(1)}(r_i) +   
\sum_{i=N_1+1}^{N} u_1^{(2)}(r_i) \right) 
\end{equation}
the one-body terms  and  
\begin{equation}
\psi_2 (\mathbf{R}) = \exp \left( \sum_{i=1}^{N_1-1} \sum_{j=i+1}^{N_1} 
u^{(11)}_2(r_{ij}) + \sum_{i=N_1+1}^{N-1} \sum_{j=i+1}^{N} 
u^{(22)}_2(r_{ij}) + \sum_{i=1}^{N_1} \sum_{j=N_1+1}^{N} 
u^{(12)}_2(r_{ij}) \right)
\end{equation}
the two-body ones.
The one-body terms $u_1^{(\alpha)}(r)$ are chosen as the analytical 
solutions (gaussian functions)  of a single particle confined by the harmonic 
trap, either 
isotropic or anisotropic. These terms are gaussian functions and depend on the specific species of the mixture. For the two-body terms associated to the interatomic dipolar potentials,
 we use the short-distance approximation of the solution for two untrapped interacting dipoles 
\cite{phdthesis},
\begin{equation}
    u_2^{(\alpha \beta)}(r_{ij})  = - 
\sqrt{\frac{2\mu_{\alpha\beta}C_{dd}^{(\alpha\beta)}}{4\pi\hbar^2}}\frac{2}{
\sqrt{r_{ij}}} \ ,
    \label{twodipoles}
\end{equation}
with $\mu_{\alpha\beta} = m_{\alpha}m_{\beta}/(m_{\alpha}+m_{\beta})$ the 
reduced 
mass of an $\alpha-\beta$ pair, and $C_{dd}^{(\alpha\beta)} = 
\mu_{0}\mu_{\alpha}\mu_{\beta}$.  Wave functions (\ref{twodipoles}) are too 
repulsive  
for intermediate and large inter-particle distances. Therefore, we match Eq. (\ref{twodipoles}) with  softer 
$u_2$ terms for $r^*_{ij}=R_{\text{cut}}^{(\alpha \beta)}$,
\begin{equation}
    u_{2}^{(\alpha\beta)}(r_{ij}^{*})= c_{2}^{(\alpha\beta)}-\frac{c_{3}^{(\alpha\beta)}}{r_{ij}^{*}}, \label{eq:two-body_wf_after}
\end{equation}
with $c_2^{(\alpha\beta)}$, $c_3^{(\alpha\beta)}$, and 
$R_{\text{cut}}^{(\alpha\beta)}$ 
constants 
for each type of $\alpha-\beta$ interaction, such that the 
wave function and its first derivative are continuous at 
$r_{ij}^{*}=R_{\text{cut}}^{(\alpha\beta)}$. This is then introduced in the VMC code 
with $R_{\text{cut}}^{(\alpha\beta)} = r_{p}^{(\alpha\beta)}A_{1}^{-1/4}$ in the 
isotropic case, with $r_{p}^{(\alpha\beta)}$ the only free variational parameters with 
respect to which the trial wave function is optimized. In the presence of an 
anisotropic harmonic potential this is adapted considering 
$r_{p}^{(\alpha\beta)} \sqrt{A_{x}^{-1/4}A_{y}^{-1/4}}$ as a more convenient 
choice for the matching distance.

\subsection{The miscibility criterion} \label{subsec:miscibility}

The miscibility in Bose-Bose mixtures is determined by the parameter $\Delta$,
\begin{equation}
    \Delta = \frac{g_{11}g_{22}}{g_{12}^2} - 1 \ , 
    \label{eq:Delta_mean-field}
\end{equation}
a criteria derived from the mean-field treatment of 
bulk mixtures \cite{cikojevic2018harmonically}. 
Equation~\ref{eq:Delta_mean-field} 
classifies the behavior of the mixtures: $\Delta>0$ signals miscibility, 
$\Delta<0$ phase separation, and $\Delta = 0$ the critical value separating both 
regimes.

In three dimensions, the  
interaction strengths in  Eq.~\ref{eq:Delta_mean-field} read
\begin{equation}
g_{\alpha,\beta} = \frac{2\pi\hbar^{2}a_{\alpha\beta}}{\mu_{\alpha\beta}} \ , 
\label{eq:strength_3D}
\end{equation}
with $\mu_{\alpha\beta} = m_{\alpha}m_{\beta}/(m_{\alpha}+m_{\beta})$ the
reduced mass and $a_{\alpha\beta}$ the $s$-wave scattering length. 
In a purely two-dimensional system,  
$g_{\alpha,\beta}$ read \cite{kim1999two}
\begin{equation}
    g_{\alpha,\beta} = 
\frac{2\pi\hbar^2}{\mu_{\alpha\beta}}\frac{1}{\abs{\ln{(n_{\alpha\beta}a_{
\alpha\beta}^{2})}}} \ , 
\label{eq:strength_2D}
\end{equation}
with $n_{\alpha}$ and $n_{\beta}$ the densities of the $\alpha$ and $\beta$
species, respectively, with $n_{\alpha\beta}$ normally defined as the 
geometrical mean $\sqrt{n_{\alpha}n_{\beta}}$. The 2D scattering lengths for a 
purely repulsive interaction ($1/r^3$) are known,  
\begin{equation}
    a_{\alpha\beta} = e^{2\gamma}\frac{2\mu_{\alpha\beta} 
C_{dd}^{(\alpha\beta)}}{4\pi\hbar^{2}} \ , 
\label{eq:scattering_length}
\end{equation}
with $\gamma = 0.5772156649...$ the Euler's constant.

It is worth noticing that, in contrast 
to Bose-Bose mixtures with contact 
interactions~\cite{cikojevic2018harmonically} where the scattering lengths 
$a_{11}$, 
$a_{22}$, and $a_{12}$ can all be changed independently, in our dipolar system 
setting $m_2/m_1$ 
and $\mu_2/\mu_1$ does not only fix $a_{11}$ and $a_{22}$, but the crossed 
scattering length $a_{12}$ as well (see Eq.~(\ref{eq:scattering_length})). 
This dependence constrains the accessible  regions of the phase-space given by 
$a_{12}/a_{22}$ 
and $a_{11}/a_{12}$, which means that certain regions that would normally be 
mapped in order to find characteristic spatial configurations, given by 
specific 
relations between the scattering lengths, are unreachable.

The miscibility criterion in 2D is  
more complex  than that in 3D (\ref{eq:strength_3D}) because of the 
explicit dependence of $g_{\alpha,\beta}$ on the densities. In a bulk 
system, this is not an issue and 
the phase-space can be determined using Eq.~(\ref{eq:Delta_mean-field}). 
However, these densities are not clearly defined in a finite system and, in 
addition, cannot be predicted exclusively from the knowledge of the external 
parameters. One could think of different ways of approximating them, such as 
considering $n_{11}$ and $n_{22}$ to be the central values of the  
density profiles $\rho_{11}(r)$ and $\rho_{22}(r)$, respectively, or their peak 
values, among other options. In the present work, we tried 
these approaches regarding the densities, and also other 
$g_{\alpha\beta}$ expressions based on the chemical potentials and the harmonic 
confinement's strength \cite{karle2019coupled,kim1999two}. In all these trials, 
we were not able to match the DMC results with the miscibility criterion  
(\ref{eq:Delta_mean-field}). We conclude that the complex 2D 
$g_{\alpha\beta}$ expressions and the uncertainties regarding the proper 
definitions of $n_{11}$, $n_{22}$ and $n_{12}$ in finite, confined systems make 
an exploration of the phase-space according to 
Eq.~(\ref{eq:Delta_mean-field}) unfeasible.

On the other hand, we observed that our results on the miscibility of the 
dipolar mixture are quite 
well determined through the definition of the adimensional parameter
\begin{equation}
\Delta = \frac{m_2}{m_1}\left(\frac{\mu_1}{\mu_2}\right)^{2} \ . 
\label{eq:delta_MC}
\end{equation}
According to this empirical parameter, we concluded that  $\Delta \simeq 1$ 
signals miscibility, with both $\Delta \gtrsim 1.1$ and $\Delta \lesssim 0.9$ 
indicating phase separation. The application of this parameter is discussed in 
the next Section.

\section{Results} 
\label{sec:results}

\subsection{Isotropically trapped mixtures} 
\label{subsec:results_iso}

We focus exclusively on systems with $N_1=N_2=100$ and a harmonic confinement of 
strength $A=0.1$ in reduced units. We do this to limit the number of free 
parameters and to focus on a system size for which a balance is found between 
the importance of interactions and the computational cost. Under these 
conditions, the central densities are of order one, always in reduced units. 

\begin{figure}[h]	
%\widefigure
\begin{center}
\includegraphics[width=10.5 cm]{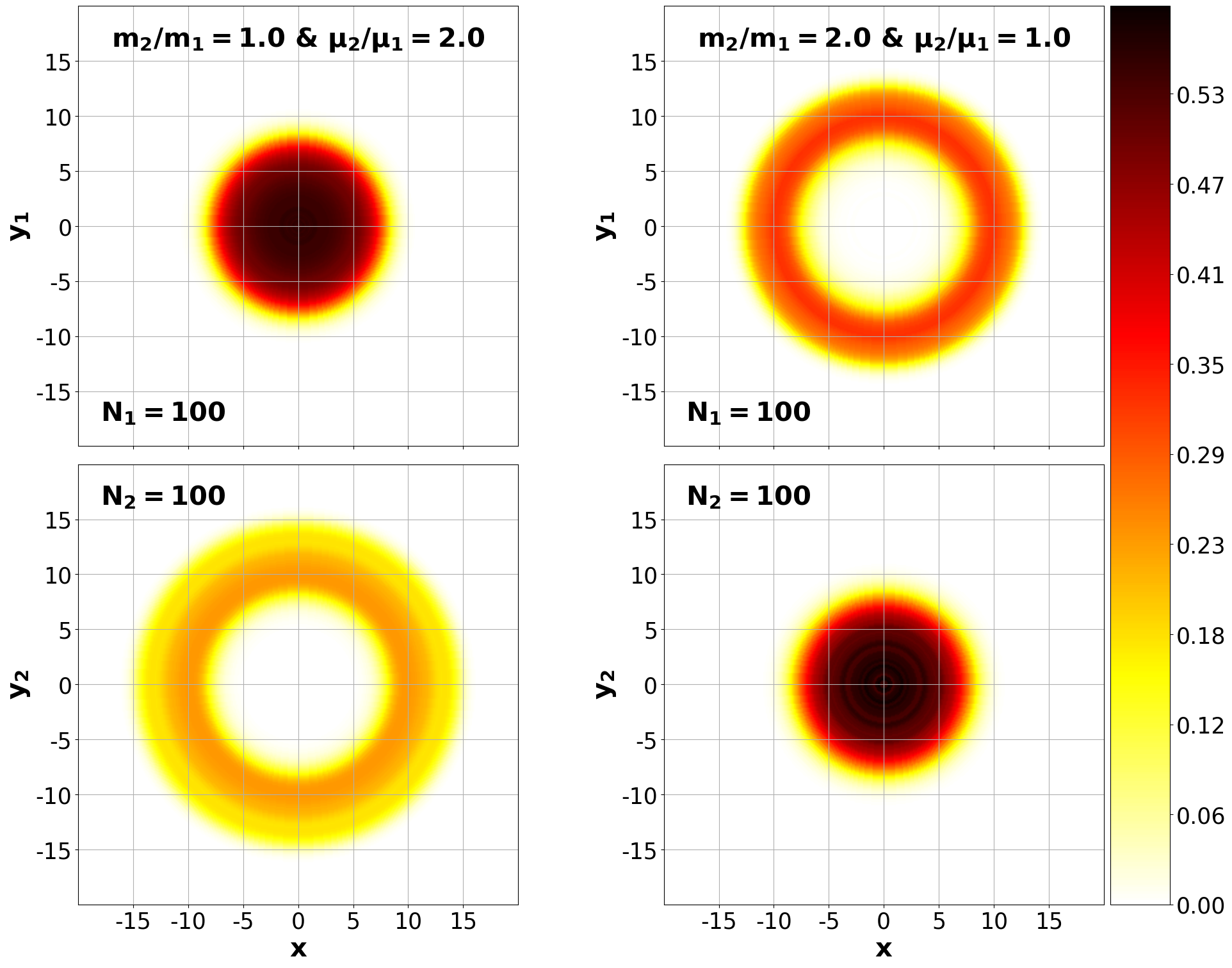}
\end{center}
\caption{Left column: pure estimators of the density for $(m_2/m_1 = 1, 
\mu_2/\mu_1 = 2)$. Right column: pure estimator of the density for $(m_2/m_1 = 
2, \mu_2/\mu_1 = 1)$.\label{fig:dmc_iso_paradigmatic_rho}}
\end{figure}  

We start with the following paradigmatic mixtures: $(m_2/m_1 = 1, \mu_2/\mu_1 
= 2)$ and $(m_2/m_1 = 2, \mu_2/\mu_1 = 1)$. These will allow to understand 
the effects of the mass and magnetic dipole moment relations between 
species before an exhaustive analysis of the phase diagram is performed. The 
pure estimators for the radial density functions $\rho(r)$ are shown in 
Fig.~\ref{fig:dmc_iso_paradigmatic_rho}. We observe that if the mass of the 
two species is the same, the particles with the larger magnetic dipole moment 
move out of the center and completely surround the other species. This is due to 
the type $2$ particles repelling each other  more 
strongly than type $1$ particles do, leading the system to a configuration where 
type $2$ particles are as separated as possible from one another and from the 
other species. The DMC simulation tells us that the most energetically 
favorable way to do this is by the second species forming a ring around the 
first one. Regarding the right panel in 
Fig.~\ref{fig:dmc_iso_paradigmatic_rho}, we see that if the magnetic dipole 
moment of the two species is the same, it is the lighter particles that now 
envelop the heavier ones. This is a direct consequence of both species being 
under the presence of the same confinement, as the lighter one presents a larger 
harmonic length.  

\begin{figure}[h]
%\widefigure
\begin{center}
\includegraphics[width=13.5 cm]{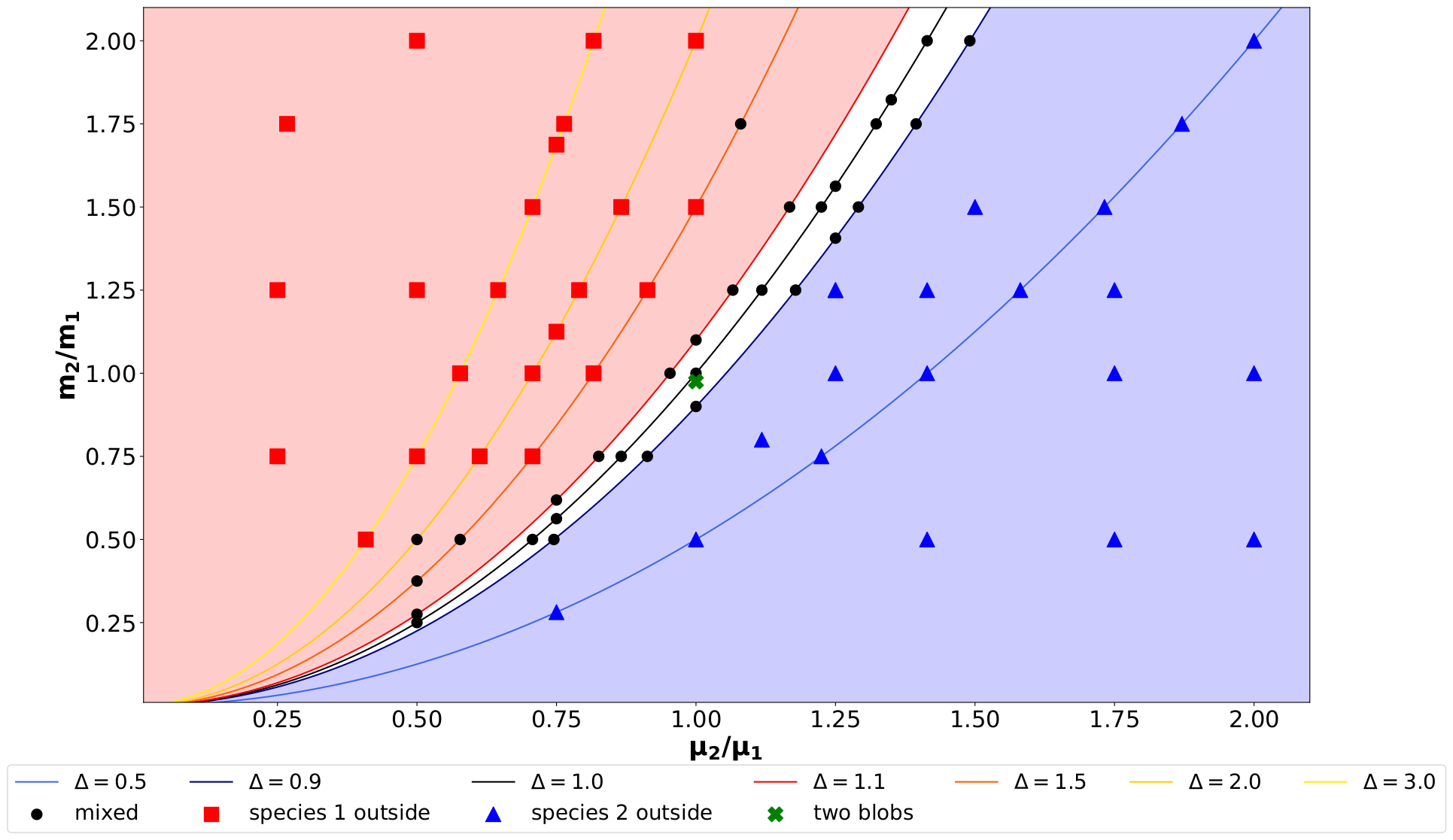}
\end{center}
\caption{Miscibility of balanced mixtures as a function of $m_2/m_1$ and
$\mu_2/\mu_1$ for $N = 200$ and $A = 0.1$. The points indicate the performed 
simulations. \label{fig:phase_diagram}}
\end{figure}

The phase diagram of the mixture is obtained by carrying out  multiple 
simulations for different $m_2/m_1$ and $\mu_2/\mu_1$ values. From the density 
profiles of both species, we determine if they are miscible or not and these 
results are compared with the empirical criterion of Eq.~(\ref{eq:delta_MC}). 
The results are shown in Figure~\ref{fig:phase_diagram}. 

As it can be seen in Fig.~\ref{fig:phase_diagram}, the criterion 
(\ref{eq:delta_MC}) matches almost perfectly with the DMC results, 
with a very narrow region of miscibility at $0.9 \lesssim \Delta \lesssim 1.1$, 
indicating the system's clear tendency to phase-separate as soon as it is 
possible due to the repulsiveness of the DDIs. In addition, every system that 
phase separates does so via one species leaving the center and surrounding the 
other one. The farther $\Delta$ is from $1.0$, the more completely it surrounds 
the other, leaving the center entirely for values of $\Delta = 3.0$ and $\Delta 
= 0.5$, with complete separation already happening for values closer to $\Delta 
= 1.0$ for specific $m_2/m_1$ and $\mu_2/\mu_1$ relations. In addition, we 
noticed that for the mixtures where the spatial configuration is driven by 
$\mu_2/\mu_1$ the phase-separation is clearer, with one species abandoning the 
center entirely for values closer to $\Delta = 1.0$. Moreover, let us remark how 
$\Delta$ can also be used to predict which particles leave the core, as for 
$\Delta > 1.1$ species 1 does it, while for $\Delta < 0.9$ it is species 2 that 
occupies the external shell. 

There is one exception in Fig.~\ref{fig:phase_diagram} corresponding to the 
particular values $\mu_2/\mu_1 = 1.0$ and $m_2/m_1 = 
0.975$ (green point). In this case, the system does not entirely mix, but 
neither does it phase-separate in the usual form. As reported in 
Ref.~\cite{cikojevic2018harmonically} for 3D  
harmonically trapped Bose-Bose mixtures with contact interactions, in 
certain cases the system clearly phase-separates forming a ``two-blobs" 
configuration, where each species occupies a  semi-circumference. 
 However, in our case, 
under an isotropic confinement the system does not phase-separate completely, 
and seems to only hint at such a configuration. This seems to be an exception, 
as we were unable to find any other case where this happened, and so we leave 
the analysis of this particular system to the next subsection, where we will 
discuss how for a given deformation of the trap, the ``two-blobs" 
configuration can be observed before it disappears for stronger compressions.

\subsection{Anisotropically trapped mixtures} 
\label{subsec:results_aniso}

In this section, we focus on systems with $N_1=N_2=100$ particles and analyze 
progressively stronger sets of anisotropic harmonic confinements with $A_x = 
0.1$ fixed and $A_y$ changing from $0.5$ to $2.5$.  We do 
not want $A_y$ to be too large as we do not wish to enter the one-dimensional 
regime;  we are interested 
in the effects of deforming the system, not changing its behavior completely. 
We study three mixtures given by $(m_2/m_1 = 1.0,\mu_2/\mu_1 = 2.0)$, $(m_2/m_1 
= 2.0,\mu_2/\mu_1 = 1.0)$, and $(m_2/m_1 = 0.975,\mu_2/\mu_1 = 1.0)$, which 
hereinafter we label as  $A$, $B$ and $C$, respectively. Mixture 
$A$  will 
allow us to isolate the effects of changing the magnetic dipole moment of one 
species and mixture $B$ will do the same with the mass relation. Finally, 
mixture $C$, as explained in Section~\ref{subsec:results_iso}, seemed to hint at 
a two-blobs configuration under an isotropic confinement, and so we will study 
how it evolves as it is compressed.

\begin{figure}[t]
%\widefigure
\begin{center}
\includegraphics[width=13.5 cm]{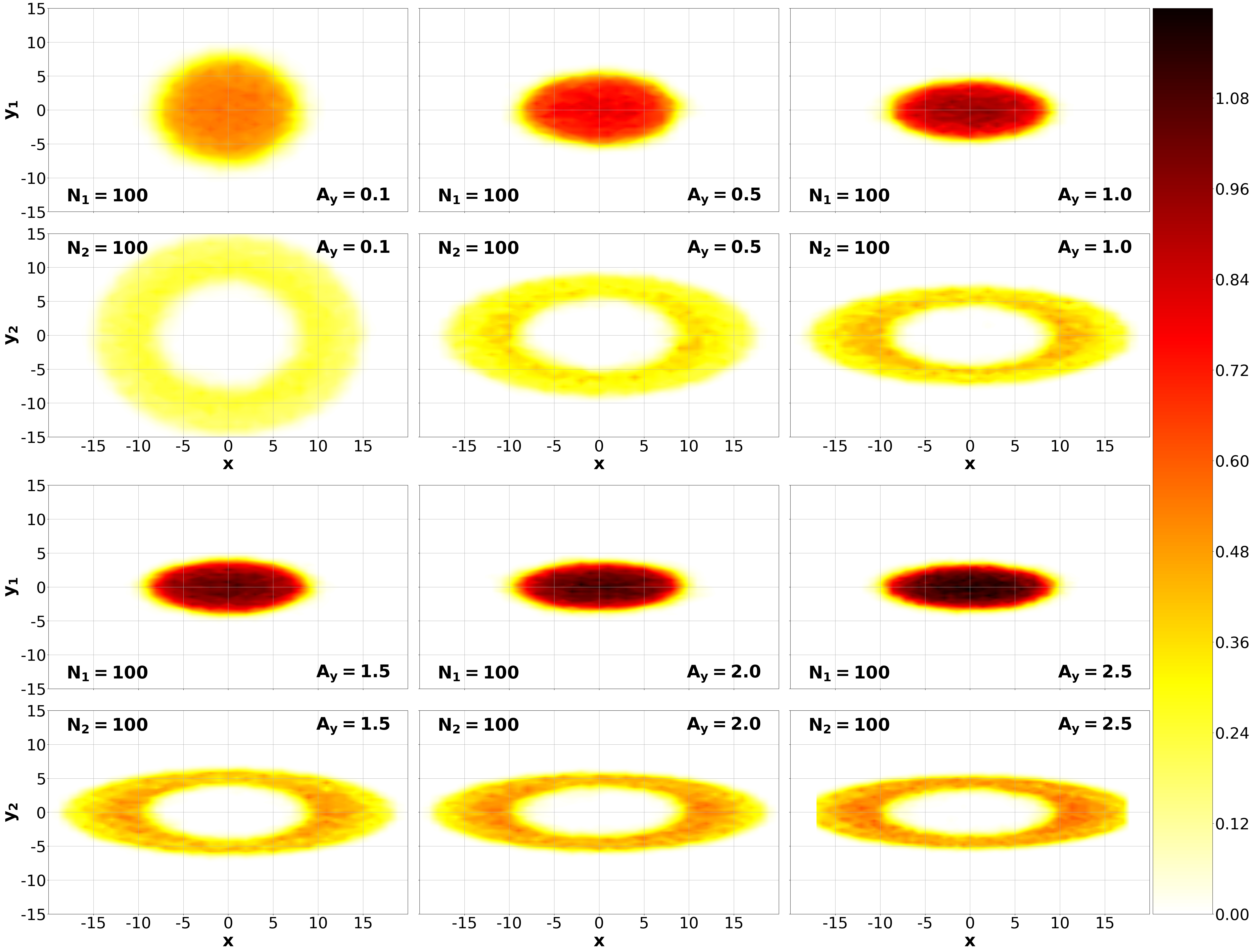}
\end{center}
\caption{Mixture A. First and second rows, from left to right: pure density 
profiles for $A_y = 0.1$, $A_y = 0.5$ and $A_y = 1.0$, respectively. Third and 
fourth rows, from left to right: density profiles for $A_y = 1.5$, $A_y = 2.0$ 
and $A_y = 2.5$, respectively. In all cases $A_x = 0.1$.}
\label{fig:dmc_aniso_balanced_mixture_A}
\end{figure}

The density profiles of mixture $A$, obtained with pure estimators, are shown 
in Fig.~\ref{fig:dmc_aniso_balanced_mixture_A}. In this case,
the type $2$ particles intra-species interactions, and the cross 
interactions with particles $1$, are so repulsive that no matter 
how 
much we compress the system in the $y$-direction, particles $2$ always 
leave the center forming a ring surrounding the species with the smaller 
magnetic dipole moment. Therefore, if the DDIs are stronger enough 
for one species compared to the other, the ring configuration is 
maintained throughout the compression.

\begin{figure}[t]
%\widefigure    
\begin{center}
\includegraphics[width=13.5 cm]{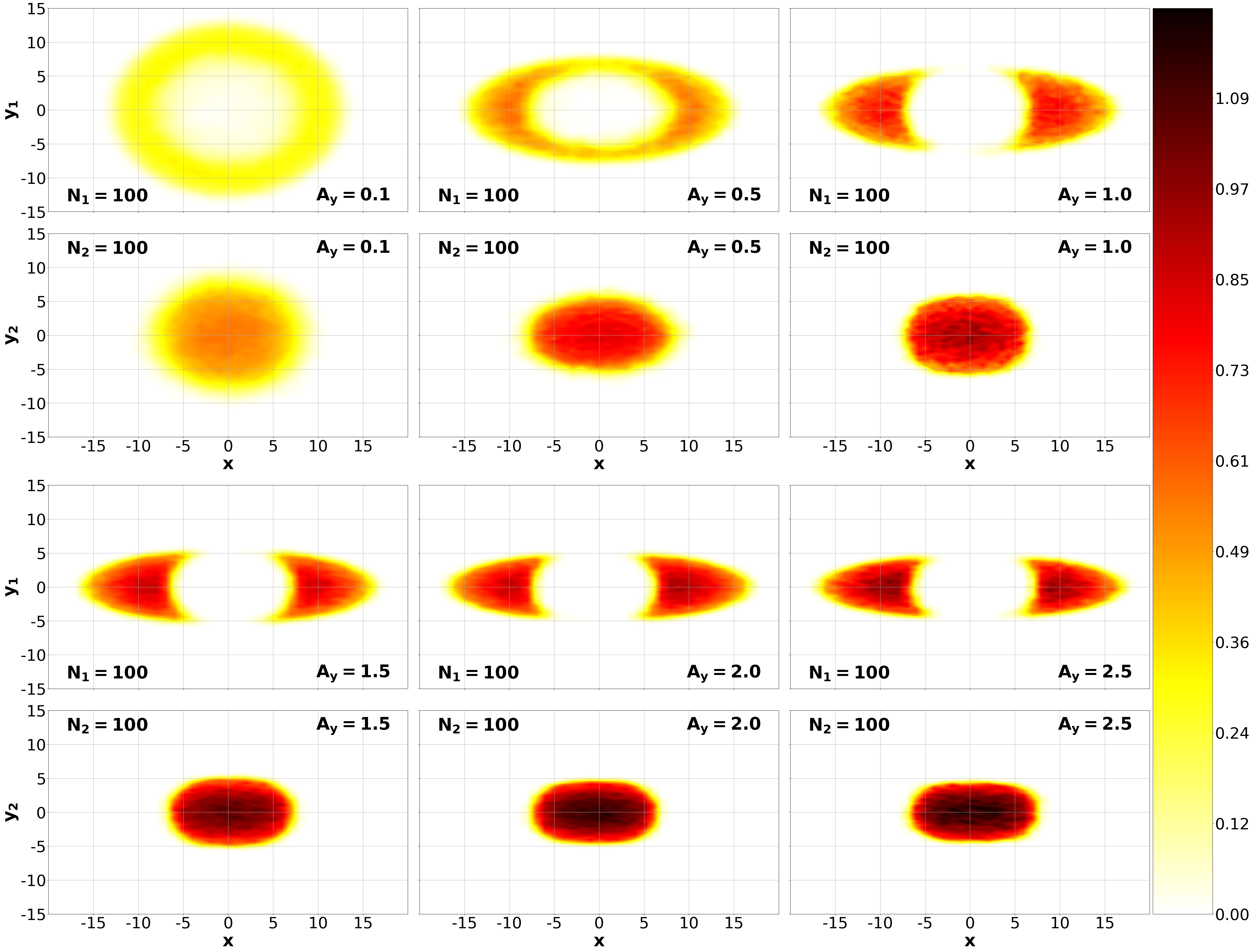}
\end{center}
\caption{Mixture B. First and second rows, from left to right: pure density 
profiles for $A_y = 0.1$, $A_y = 0.5$ and $A_y = 1.0$, respectively. Third and 
fourth rows, from left to right: density profiles for $A_y = 1.5$, $A_y = 2.0$ 
and $A_y = 2.5$, respectively. In all cases $A_x = 0.1$.}
    \label{fig:dmc_aniso_balanced_mixture_B}
\end{figure}

The density profiles for mixture $B$ are displayed in 
Fig,~\ref{fig:dmc_aniso_balanced_mixture_B}. We can see that in this case 
the configuration is not maintained throughout the deformation. If the 
compression is small (see $A_y = 0.5$) the ring is still present, as it is 
still energetically favorable for the lighter species to leave the center, 
surrounding the others. However, as the compression progresses, the equilibrium 
between the DDIs and the harmonic confinement for  particles 1 cannot be 
reached with the ring configuration.  Particles 2 now occupy the 
center in such a way that the second species cannot keep enough distance 
between 
themselves and the other species at small and intermediate $x$ values for it to 
be energetically favorable. As a result, particles 1  form 
"wings" surrounding particles 2.

\begin{figure}[t]
%\widefigure
\begin{center}
\includegraphics[width=13.5 cm]{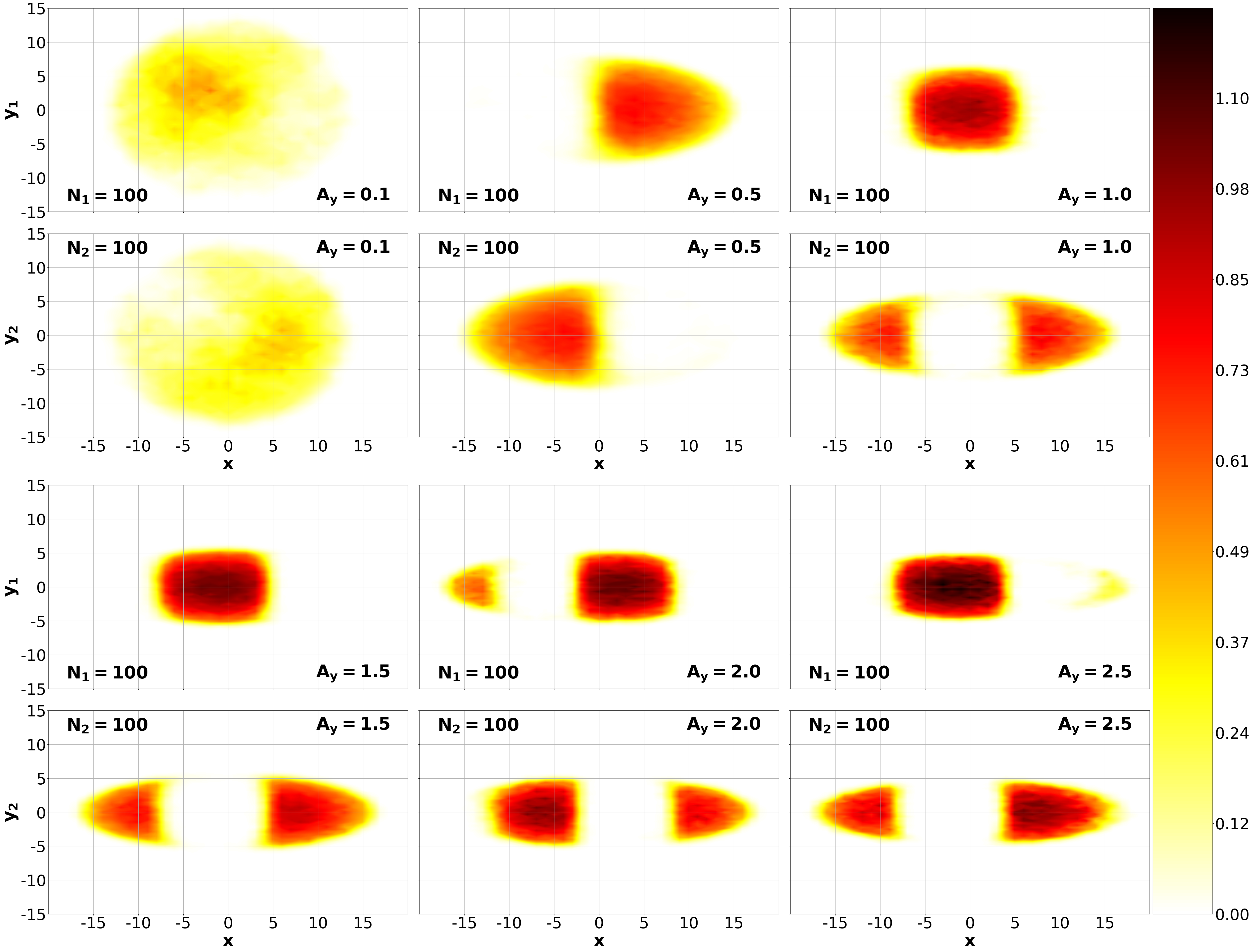}
\end{center}
\caption{Mixture C. First and second rows, from left to right: pure density 
profiles for $A_y = 0.1$, $A_y = 0.5$ and $A_y = 1.0$, respectively. Third and 
fourth rows, from left to right: density profiles for $A_y = 1.5$, $A_y = 2.0$ 
and $A_y = 2.5$, respectively. In all cases $A_x = 0.1$.}
\label{fig:dmc_aniso_balanced_mixture_C}
\end{figure}

Finally, mixture $C$'s density profiles are shown in 
Fig.~\ref{fig:dmc_aniso_balanced_mixture_C}. In this case, we 
start with the system hinting at a two-blobs configuration, which gets 
completely clear and defined for $A_y = 0.5$. However, it disappears for 
stronger deformations, and gives way the wing configuration described for 
mixture $B$. This system, as far as we can tell from the analyzed mixtures, is 
an exception, as we have been unable to find another system presenting these 
symmetric semi-ellipses. We suspect that this type of configuration may only 
appear at very particular $m_2/m_1$ and $\mu_2/\mu_1$ relations for which $1-2$ 
interactions are the most repulsive and $1-1$ and $2-2$ interactions are of 
similar strength. In this situation, the mixture could form symmetric 
distributions for each species, allowing for large interparticle distances.

Any other mixed isotropic configuration that we have 
analyzed turns into a phase-separated system as it is compressed. As we saw in 
Fig.~\ref{fig:phase_diagram}, it required very specific mass and magnetic 
dipole relations between the two species for the system to be in a mixed state. 
Therefore, those very specific conditions move and maybe even disappear as the 
system is compressed and the DDIs become more repulsive. It may be possible, 
though, that different regions of existence for the miscible state exist for 
these anisotropic confinements. It is, however, beyond the scope of this work to 
probe the full phase-space, as it was done for the isotropic case,  
Fig.~\ref{fig:phase_diagram}.

\subsection{Erbium-Dysprosium mixture} 
\label{subsec:results_erdy}

Throughout this work we have studied a generic set of systems, given by changes 
in the mass and magnetic dipole moment relations between species and the 
deformation of the harmonic confinement applied to it. In this Section, we  
perform a study of these effects for an Erbium-Dysprosium (Er-Dy henceforth) 
system, as it is the first Bose-Bose dipolar mixture that has been realized 
experimentally \cite{Ravensbergen,trautmann2018dipolar,Politi} and offers us a 
perfect opportunity 
to predict potentially observable configurations. Both Er and Dy are part of the 
magnetic rare-earth species group, and we are going to focus on the case of a 
${}^{166}$Er-${}^{164}$Dy mixture, with ${}^{166}$Er having an atomic mass of 
$165.930293$~uma and a magnetic dipole moment 
of $7\mu_{B}$, and ${}^{164}$Dy having a mass of 
$163.929175$ and a magnetic dipole moment of $10\mu_{B}$.  For 
convenience, we define the Er atoms as the type 1 species and the Dy ones as 
type 2, so that the mass and magnetic dipole moment relations characterizing the 
system are $m_2/m_1 \simeq 0.9897$ and $\mu_2/\mu_1  \simeq 1.4285$, 
respectively.

\begin{figure}[h]
%\widefigure
\begin{center}
\includegraphics[width=10.5 cm]{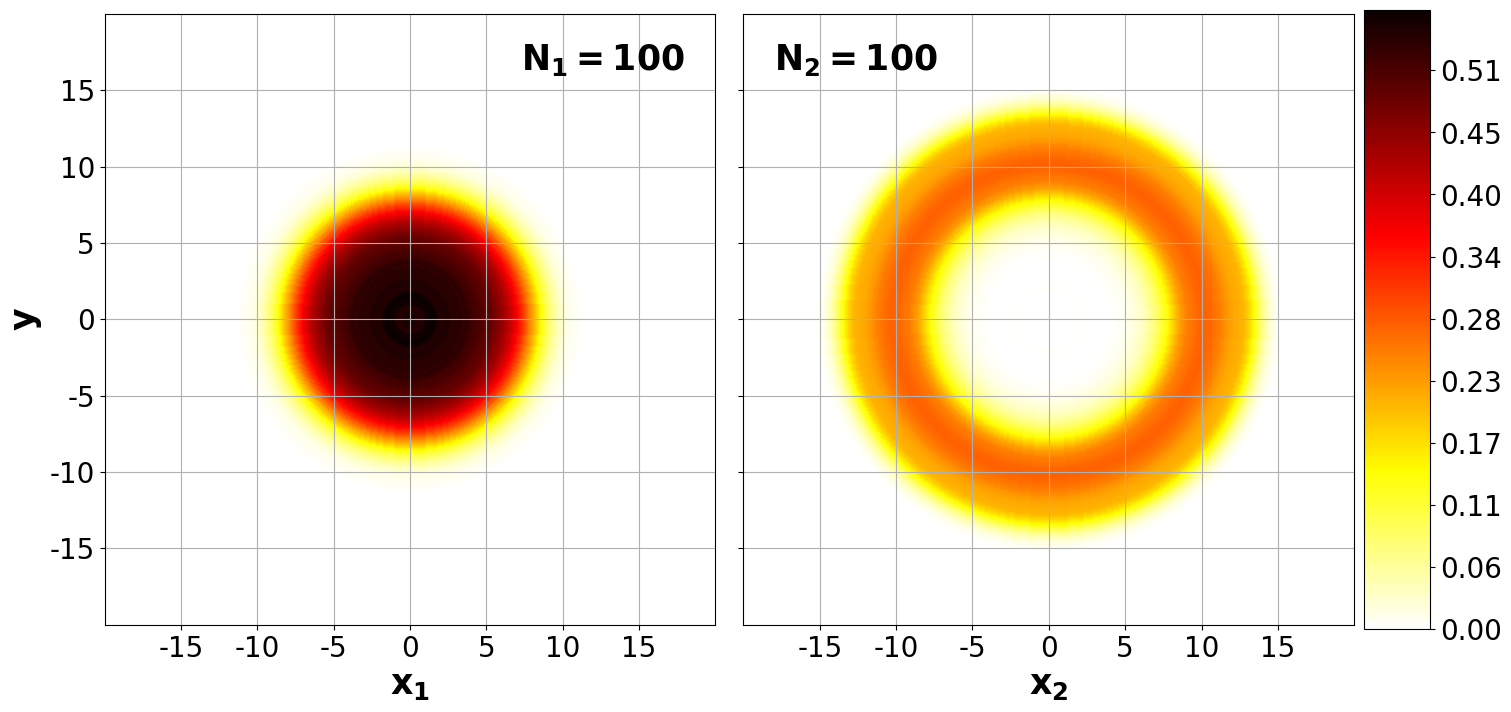}
\end{center}
\caption{$\rho_{pure}(x,y)$ for an Er-Dy balanced mixture under an isotropic 
confinement of value $A = 0.1$ in reduced dipolar units.}
\label{fig:dmc_iso_balanced_ErDy}
\end{figure}

We start with a balanced mixture of Er-Dy atoms, with $N_1 = N_2 = 100$ 
particles 
each, under the presence of an isotropic harmonic confinement of strength $A = 
0.1$. The pure density profiles for this case are shown in 
Fig.~\ref{fig:dmc_iso_balanced_ErDy}. 
As it can be clearly seen, under these conditions the system phase-separates, 
with the Dy atoms leaving the center and surrounding the Er particles. In this 
case $m_2/m_1 \approx 1$, meaning the mass difference between species is not a 
key factor, and the inmiscibility is due to the Dy atoms having a 
significantly larger magnetic dipole moment, which forces them out of the center 
due to the stronger repulsive DDI interactions between themselves and the Er 
atoms.

\begin{figure}[h]
%\widefigure
\begin{center}
\includegraphics[width= 13.5 cm]{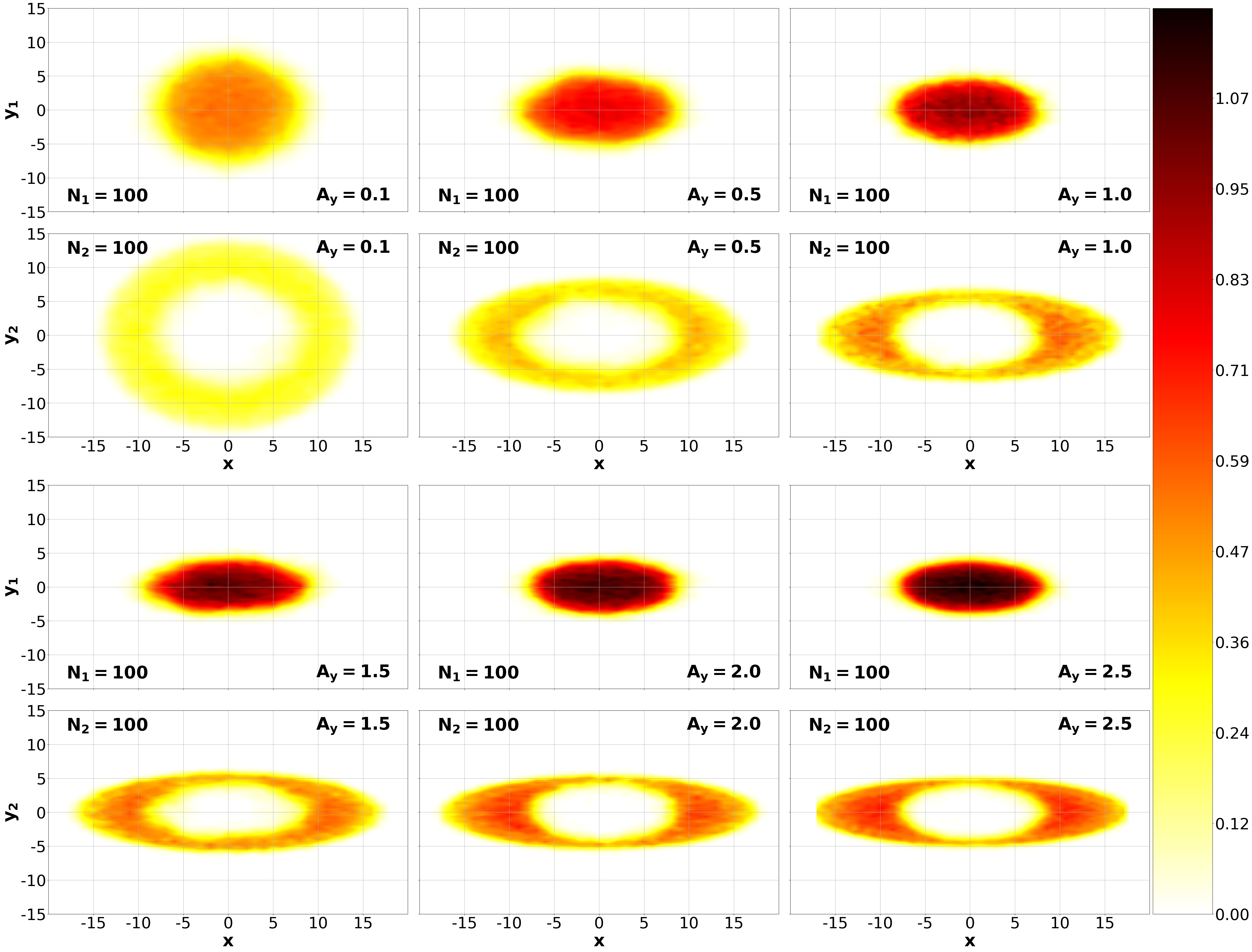}
\end{center}
\caption{First and second rows, from left to right: pure density profiles of 
the  Er-Dy mixture for $A_y = 0.1$, $A_y = 0.5$ and $A_y = 1.0$, respectively. 
Third and fourth row, from left to right: density profiles for $A_y = 1.5$, $A_y 
= 2.0$ and $A_y = 2.5$, respectively. In all cases $A_x = 0.1$.}
\label{fig:dmc_aniso_balanced_ErDy}
\end{figure}

We  expose the mixture to 
progressively stronger potentials in the $y$-direction, in the same way as in 
Sec.~\ref{subsec:results_iso}. The pure density profiles are shown in 
Fig.~\ref{fig:dmc_aniso_balanced_ErDy}. In this case, due to the mass relation 
between species essentially not playing any role, it is the DDIs that are 
integral to the obtained spatial configurations. These are so repulsive that, 
again, the ring configuration is the most favorable one, regardless of the 
applied deformation in the $y$-direction. However, as we can see from the $A_y = 
2.5$ result, it seems that pushing the system further the 
wing configuration could eventually appear.   

\section{Discussion} 
\label{sec:Discussion}

 We have presented a thorough analysis of dipolar binary 
Bose-Bose mixtures at zero temperature in two dimensions, confined harmonically 
in the $xy$ plane. The results have been obtained by a combination of two 
quantum Monte Carlo methods: variational Monte Carlo, for the 
optimization of the trial wave function, and diffusion Monte Carlo to get a 
statistically exact solution by means of solving the Schrödinger equation in 
imaginary time. With this, we have been able to map the miscibility of balanced 
mixtures under an isotropic harmonic confinement, studied anisotropic 
confinements by progressively applying a stronger trapping in the $y$ direction, 
and performed a study of the Erbium-Dysprosium mixture, the first 
dipolar Bose-Bose mixture realized experimentally.

In Sec.~\ref{subsec:results_iso}, we performed the complete analysis of 
mixtures with a total number of particles of $N=200$ under an isotropic harmonic 
confinement of strength $A = 0.1$.  We saw how the 
masses being equal, the species with the larger magnetic dipole moment left the 
center and surrounded the other ones in a ring configuration, while  $\mu_1 = 
\mu_2$ lead to the lighter particles leaving the core as a consequence of having 
a larger harmonic characteristic length. We  studied the 
miscibility of the system as a function of $m_2/m_1$ and $\mu_2/\mu_1$ with the 
help of a dimensionless parameter $\Delta$ (see Equation~\ref{eq:delta_MC} and 
Figure~\ref{fig:phase_diagram}), and concluded that the miscibility region of 
the system is very narrow, as it only occurs around $0.9 \lesssim \Delta 
\lesssim 1.1$, otherwise the system phase-separates in two different ways: for 
$\Delta >1$ species 1 leaves the center, while for $\Delta < 1$ it is species 2 
that abandons the core. Every system system studied that phase-separated did so 
by means of this 'ring' configuration in which one species occupies the core, 
and the other encircles it, as due to the repulsive DDIs this allows for the 
maximum distance between particles of both the same and different species.

In Sec.~\ref{subsec:results_aniso}, we focused on analyzing three mixtures 
($A$, $B$, and $C$) given  
under progressively stronger confinements in the $y$ direction, keeping the $x$ 
component constant. We concluded that for demixing driven by the the magnetic 
dipole moments the ring configuration remained the stable one in the 
deformation regime studied, while for mass-driven phase-separation under strong 
enough compression, the ring configuration turns to a  
wing structure, in which the lighter species is pushed to the edges of the 
system in the $x$ axis. 
In addition, we studied a very 
particular case, an exception from Section~\ref{subsec:results_iso}, the 
$(m_2/m_1 = 0.975, \mu_2/\mu_1 = 1)$ mixture, which under an isotropic 
confinement hinted at a 'two-blobs' configuration, in which two symmetrical 
semi-circumferences are formed by each species 
separately \cite{cikojevic2018harmonically} for a simpler contact interaction 
in a three-dimensional case. This system, under a moderately strong deformation 
lead to a very defined two-blobs configuration, which transformed into the 
wing one under stronger compressions. We tried to find other systems with this 
behavior, but we were ultimately unable to discover them.

To conclude, we studied the Er-Dy mixture, the first experimentally 
realized one\cite{Ravensbergen,trautmann2018dipolar,Politi}, in order to make 
predictions on its miscibility. We 
observe how under an isotropic confinement the mixture phase-separated, with the 
Dy atoms leaving the center due to their greater magnetic dipole moments. We 
then analyzed the same mixture under progressively stronger confinements in the 
$y$ direction and observed how the ring configuration held for every single 
strength that we applied. Recent experiments on Er-Dy mixtures in three 
dimensions show that the system prefers to be phase separated due to an overall 
repulsion between the two species. Moreover, the gravitational sag between both 
species, due to their different masses, makes difficult to get a full overlap 
between both species \cite{Politi}. It would be interesting to deform 
the trapping potential and approach the 2D limit, with the magnetic moments 
perpendicular to the plane, because then the gravitational effect would be 
reduced and thus the possible miscibility between both components could be 
better resolved.    

\section{Acknowledgments}

This work has been supported by AEI (Spain) under grant No. 
PID2020-113565GB-C21.   We also acknowledge financial support from Secretaria 
d'Universitats i Recerca del Departament d'Empresa i Coneixement de la 
Generalitat de Catalunya, co-funded by the European Union Regional Development 
Fund within the ERDF Operational Program of Catalunya (project QuantumCat, ref. 
001-P-001644).

\bibliography{refs.bib}

\end{document}